\documentclass{vgtc}
\usepackage{amsmath,amsfonts}
\usepackage{algorithmic}
\usepackage{algorithm}
\usepackage{booktabs}
\usepackage[table,svgnames]{xcolor}
\usepackage{colortbl}
\usepackage{array}
\usepackage{textcomp}
\usepackage{stfloats}
\usepackage{url}
\usepackage{verbatim}
\usepackage{graphicx}
\usepackage{cite}
\usepackage{comment}
\usepackage{soul}
\usepackage{transparent}
\usepackage{tcolorbox}
\usepackage{setspace}
\usepackage{tikz}
\usepackage{amssymb}
\usepackage{fancyhdr}
\usetikzlibrary{calc}
\usepackage{wrapfig}
\usepackage{endnotes}
\usepackage{multirow}
\definecolor{OliveGreen}{RGB}{183, 207, 178}
\definecolor{DarkYellow}{HTML}{DEA601}
\definecolor{DarkOrange}{HTML}{ED7D31}
\definecolor{DarkBlue}{HTML}{4472C4}
\usepackage{xspace}

\newcommand*\rot{\rotatebox{90}}
\newcommand{\nonterm}[1]{\textbf{\textsf{#1}}}
\newcommand{\literal}[1]{\texttt{#1}}
\newcommand{\typeval}[1]{\textit{#1}}
\newcommand{\dictkey}[1]{\texttt{\textcolor{black}{#1}}}

\definecolor{tsblue}{HTML}{2C3E50}   %
\definecolor{tsteal}{HTML}{2F855A}   %
\definecolor{tsorange}{HTML}{C05621} %

\renewcommand{\nonterm}[1]{\textbf{\textcolor{tsblue}{\textsf{#1}}}}
\renewcommand{\literal}[1]{\textcolor{tsteal}{\texttt{#1}}}
\renewcommand{\typeval}[1]{\textcolor{tsorange}{\textit{#1}}}

\newcommand{\optdefeq}{\stackrel{\text{opt}}{\mathrel{:=}}}
\newcommand{\defeq}{\mathrel{:=}}
\newcommand{\alt}{\;|\;}
\newcommand{\sep}{,\;}
\newcommand{\seq}[1]{\langle{}#1\rangle{}}

\usepackage[compact]{titlesec}
\titlespacing{\section}{0pt}{1ex}{0ex}
\titlespacing{\subsection}{0pt}{1ex}{0ex}

\makeatletter
\renewcommand\section{\@startsection{section}{1}{\z@}%
  {-2ex \@plus -1ex \@minus -.2ex}%
  {0.8ex \@plus .2ex}%
  {\@afterindentfalse\reset@font\normalsize\sffamily\bfseries\scshape\vgtc@sectionfont}}

\renewcommand\subsection{\@startsection{subsection}{2}{\z@}%
  {-1.8ex\@plus -1ex \@minus -.2ex}%
  {0.8ex \@plus .2ex}%
  {\@afterindentfalse\reset@font\normalsize\sffamily\bfseries\vgtc@sectionfont}}

\renewcommand\subsubsection{\@startsection{subsubsection}{3}{\z@}%
  {-1.8ex\@plus -1ex \@minus -.2ex}%
  {0.8ex \@plus .2ex}%
  {\@afterindentfalse\reset@font\sffamily\normalsize\vgtc@sectionfont}}
\makeatother

\newcommand{\etal}{et al.\xspace}
\newcommand{\etals}{\etal{}'s}
\newcommand{\ie}{{i.e.,}}
\newcommand{\eg}{{e.g.,}}

\newcommand{\cf}{{cf.}}

\newcommand{\figref}[1]{\hyperref[#1]{Fig.~\ref*{#1}}}
\newcommand{\secref}[1]{\hyperref[#1]{Sec.~\ref*{#1}}}

\newcommand{\parahead}[1]
{%
    \vspace{0.07in}%
    \noindent%
    \textbf{\textit{#1.}}%
}

\definecolor{quoteColor}{HTML}{C46299}

\usepackage[normalem]{ulem}
\definecolor{linkColor}{HTML}{257E98}
\setuldepth{Berlin}
\newcommand\asLink[2]{\textcolor{linkColor}{\small\href{#1}{\ul{#2}}}}

\renewcommand{\paragraph}[1]{\textbf{#1:}}

\definecolor{indvHigh}{HTML}{9BB2E0}
\definecolor{indvMed}{HTML}{C8D6ED}
\definecolor{indvLow}{HTML}{ECF2F9}

\definecolor{grpHigh}{HTML}{B6E09B}
\definecolor{grpMed}{HTML}{D6EDC7}
\definecolor{grpLow}{HTML}{F2F9ED}

\definecolor{ensTaskHigh}{HTML}{B86029}
\definecolor{ensTaskMed}{HTML}{EAB38B}
\definecolor{ensTaskMedLow}{HTML}{EFCAB0}
\definecolor{ensTaskLow}{HTML}{F8E4D8}

\definecolor{ensTotHigh}{HTML}{B7912F}
\definecolor{ensTotMed}{HTML}{F5C242}
\definecolor{ensTotMedLow}{HTML}{F8DA78}
\definecolor{ensTotLow}{HTML}{FDF3D0}

\definecolor{dsHigh}{HTML}{2F5597}
\definecolor{dsMed}{HTML}{8FAADC}
\definecolor{dsLow}{HTML}{B4C7E7}

\onlineid{1199}

\vgtccategory{Research}

\vgtcpapertype{Theoretical \& Empirical}

\title{AnnoGram: An Annotative Grammar of Graphics Extension}

\author{%
  \authororcid{Md Dilshadur Rahman}{0009-0008-5467-615X},
  \authororcid{Md Rahat-uz- Zaman}{0000-0001-6728-7569}, 
  \authororcid{Andrew McNutt} {0000-0001-8255-4258},
  and \authororcid{Paul Rosen} {0000-0002-0873-9518}
}

\authorfooter{
  \item
  	Josiah Carberry is with Brown University.
  	E-mail: jcarberry@example.com
  \item
  	Ed Grimley is with Grimley Widgets, Inc.
  	E-mail: ed.grimley@example.com.

  \item Martha Stewart is with Martha Stewart Enterprises at Microsoft
  Research.
  	E-mail: martha.stewart@example.com.
}

\abstract{Annotations are central to effective data communication, yet most visualization tools treat them as secondary constructs---manually defined, difficult to reuse, and loosely coupled to the underlying visualization grammar. We propose a declarative extension to Wilkinson's Grammar of Graphics that reifies annotations as first-class design elements, enabling structured specification of annotation targets, types, and positioning strategies. To demonstrate the utility of our approach, we develop a prototype extension called Vega-Lite Annotation. Through comparison with eight existing tools, we show that our approach enhances expressiveness, reduces authoring effort, and enables portable, semantically integrated annotation workflows.

}


\teaser{
  \centering
  \includegraphics[width=0.97\linewidth, alt={}]{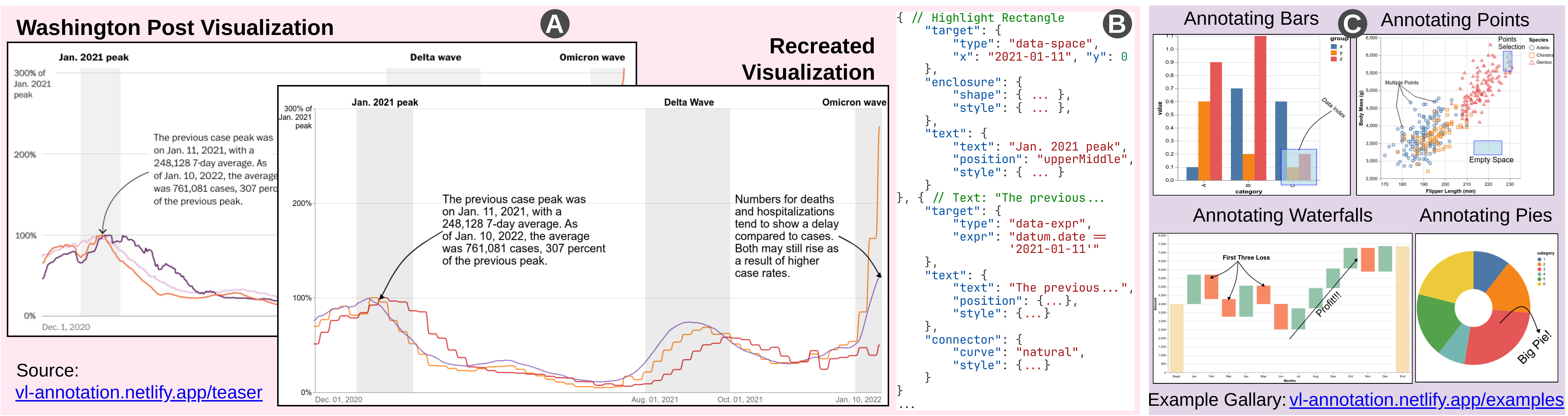}
  \caption{%
  (A)~An annotated Washington Post visualization~\cite{tan2022omicron} recreated using a proof-of-concept implementation of our AnnoGram extension to the Grammar of Graphics, (B) is shown alongside an excerpt of the specification. 
  (C)~the implementation supports diverse annotation types (text, connector, enclosure, indicator, etc.) across various chart types. 
  }
  \label{fig:teaser}
}

\graphicspath{{figs/}{figures/}{pictures/}{images/}{./}} %

\usepackage{tabu}                      %
\usepackage{booktabs}                  %
\usepackage{lipsum}                    %
\usepackage{mwe}                       %
\usepackage{wasysym}

\usepackage{svg}

\usepackage{mathptmx}                  %

\begin{document}

\maketitle

\section{Introduction}
\label{sec.intro}
Annotations are textual and graphical elements embedded in visualizations to draw attention, highlight salient features, provide context, and guide interpretation~\cite{munzner2014visualization, rahman2024qualitative, rahman2024exploring}. They improve memorability, comprehension, and recall~\cite{cedilnik2000procedural, borkin2013makes, borkin2015beyond, bateman2010useful}, and help externalize insights during analysis~\cite{Kang2014characterizing, shrinivasan2009connecting, lin2022data}. Annotations are employed in visualizations across numerous domains, including scientific research, business analytics, healthcare, education, journalism, and public policy. Despite their ubiquity, current programmatic tools do not treat annotations as first-class elements, making annotations difficult to reuse across different designs and datasets and impeding their computational analysis (such as for recommendation).

Most visualization platforms treat annotations as secondary overlays, offering only ad hoc mechanisms loosely connected to underlying data and encodings. For example, Vega-Lite~\cite{satyanarayan2016vega} lacks a dedicated annotation abstraction; users must instead define separate data sources and specify annotation marks using basic graphical primitives such as points, lines, and text. This approach is verbose, error-prone, and complicates the maintenance and updating of annotations as the data or design evolves. Similarly, D3.js~\cite{bostock2011d3} affords low-level graphical control but requires users to imperatively construct, position, and manage annotations, tightly binding them to specific chart implementations and reducing reusability. Although tools such as d3-annotation offer reusable task-specific components, their non-declarative design limits flexibility and generalization across visualization types. The situation is similar for other programmatic tools. 
As a result, many practitioners export charts to tools such as PowerPoint to add annotations manually~\cite{parsons2021understanding, bigelow2016iterating}, severing links to data, removing interactivity, and limiting portability and automation.

To address these challenges, we propose AnnoGram, an extension to Wilkinson's Grammar of Graphics~\cite{wilkinson2011grammar} (GoG) that elevates annotation to a first-class design element---decomplecting it from GoG's notion of guides and secondary datasets, and making it a sibling to scales and geometries.
Our extension 
supports declarative specification of annotation types, targets, positioning strategies, and styles. This support reduces manual positioning, enables data-driven placement, and promotes reusability, allowing annotations to transfer across charts as needs evolve.
As a demonstration, we construct a proof-of-concept implementation
that extends Vega-Lite~\cite{satyanarayan2016vega}. By extending this DSL, rather than re-implementing it, we can reuse the full Vega rendering pipeline while still making use of it as a GoG petri dish  (echoing Zong, Pollock, \etal{}~\cite{zong2022animated}). 
We evaluate the extension using a heuristic-based comparison with existing tools. 

Through this work, we show that annotations can be first-class components of visualization grammars that enabling consistent, reusable workflows.

\section{Related Work}

Our work extends prior work on tool support for visualization annotations as well as on Grammar of Graphics~\cite{wilkinson2011grammar}. 
GoG provides an expressive conceptual framework for describing a wide array of statistical graphics and has been used to structure tools like ggplot2~\cite{wickham2011ggplot2} and Vega-Lite~\cite{satyanarayan2016vega}.
The central observation in this model is that specifying charts via their types (\eg{} scatterplots or bar charts) complects different ideas (\eg{} notions of scale and encoding) that, when reified, support expressive construction of a broad range of charts, both novel and familiar. 
Yet, this model is not perfect and omits important structures for some tasks.
For instance, accessibility~\cite{mcnutt2022no} or uncertainty~\cite{pu2020probabilistic}, or as we consider, annotation. 
Notably, GoG can support each of these tasks in an unmodified form (\cf\ Wilkinson's uncertainty chapter~\cite{wilkinson2011grammar}); however, it encounters the so-called \textit{Turing tarpit}, a situation in programming language design in which ``everything is possible but nothing of interest is easy''~\cite{perlis1982special}. 
This situation is particularly evident for annotation, where current GoG-based tools require verbose and indirect encodings. We develop an extension that privileges a specific task of interest (annotation in our case) that, based on usage\cite{rahman2024survey}, is believed to be important.

Recent efforts have extended Vega-Lite’s declarative grammar to support underrepresented design concerns---an approach our work follows for annotations. 
For instance, Cicero~\cite{kim2022cicero} is a responsive design grammar that modifies Vega-Lite to enable layout adaptation across devices. 
Animated Vega-Lite~\cite{zong2022animated} extends the Vega-Lite's compiler with a temporal encoding channel to support animation, treating time as a first-class visual variable. 
Tactile Vega-Lite~\cite{pineros2025tactile} augments the grammar with tactile-specific abstractions---such as braille, textures, and spatial layout controls---to support blind and low-vision readers. 
DXR~\cite{sicat2018dxr} uses an extension of Vega-Lite syntax to support authoring immersive 3D visualizations. We similarly extend the Vega-Lite grammar but focus on the underexplored dimension of annotation, reifying it as a core design element. 
In doing so, our work complements these prior extensions by addressing a different yet equally foundational aspect of visualization design.

Beyond grammar-based approaches, many tools support annotations~\cite{willett2011commentspace, heer2007voyagers, highcharts_docs, eingartner2025inflecting}, albeit in limited ways~\cite{rahman2024survey}.
For example, ChartAccent~\cite{ren2017chartaccent} maps data to marks using rules.
Contextifier~\cite{hullman2013contextifier} automatically generates annotations from text for time series charts.
Although these systems support annotations, they commonly treat them as external overlays detached from the underlying visualization model, lacking the modularity, portability, and integration commonly found in data-driven workflows.
Our work enables annotations to be specified, reasoned about, and rendered alongside core visualization constructs. Some prior work has also explored structured annotations for reasoning~\cite{zhao2016annotation} and narrative generation~\cite{bryan2016temporal}, treating annotations as integral to analysis or storytelling. 
Our work complements these by formalizing annotation structure for declarative specification and reuse.

\section{Extension Design}
\label{sec:grammar}
Our primary contribution is AnnoGram (summarized in \autoref{fig:grammar}), an extension to GoG that elevates annotations to the same level as other primary GoG components, such as geometries or scales.  
Unlike other extensions, ours is complicated by the issue that annotations require information from across the specification-assembly-display pipeline (a concept underpinning the organization of GoG~\cite{wilkinson2011grammar} itself). For instance, annotation placement can require knowledge of the pixel position of generated marks (such as to address overlap or collision) while also needing knowledge of the data (such as to select the maximum value in a scatterplot).

\begin{figure}[t]
\centering
\resizebox{\linewidth}{!}{%
\small{}%
\setlength{\tabcolsep}{10pt}%
\setlength\arraycolsep{1pt}%
$
\begin{array}{l}
\rot{\nonterm{Root}}\left\{
\begin{array}{lcl}
    \dictkey{target} & \defeq & (\literal{id} \alt \nonterm{FixedPos} \alt \nonterm{ChartPart} \alt \nonterm{DataPoint} \alt \nonterm{Axis} \alt \typeval{None})[] \\
    \dictkey{text} & \optdefeq & \nonterm{Annotation}\seq{\nonterm{TextAnn}} \\
    \dictkey{enclosure} & \optdefeq & \nonterm{Annotation}\seq{\nonterm{EnclosureAnn}} \\
    \dictkey{connector} & \optdefeq & \nonterm{Annotation}\seq{\nonterm{ConnectorAnn}} \\
    \dictkey{indicator} & \optdefeq & \nonterm{Annotation}\seq{\nonterm{IndicatorAnn}} \\
    \cdots{}
\end{array}
\right. \\\\[-1.2em]
\multicolumn{1}{l}{\underline{\textbf{\textit{Annotation Targets}}}} \\
\begin{array}{lcl}
\nonterm{DataPoint} & \defeq & \typeval{Expression} \alt \literal{index[]} \\
\nonterm{ChartPart} & \defeq & \literal{title} \alt \literal{legend} \alt \literal{subtitle} \alt \literal{caption} \alt \cdots{} \\
\nonterm{Axis} & \defeq &
\left\{
\begin{array}{lcl}
\dictkey{axis} & \defeq & \literal{x} \alt \literal{y} \sep \dictkey{parts} \defeq \literal{label} \alt \literal{tick} \alt \literal{grid} \alt \literal{tick-label} \\
\dictkey{range} & \optdefeq & \nonterm{Coord}[] \alt \typeval{Expression}
\end{array}
\right.
\end{array} \\\\[-1.2em]
\multicolumn{1}{l}{\underline{\textbf{\textit{Annotation Types}}}} \\
\begin{array}{lcl}
\nonterm{Annotation}\seq{X} & \defeq & (\literal{id} \sep \nonterm{Style} \sep X), \quad \nonterm{Style} = \cdots{} \\
\nonterm{TextAnn} & \defeq & (\typeval{string} \sep \nonterm{Position} \alt \typeval{None}) \\
\nonterm{EnclosureAnn} & \defeq & (\nonterm{Rect} \alt \nonterm{Ellipse} \alt \literal{SVGShapePath} \alt \cdots{} \sep \nonterm{Position} \alt \typeval{None}) \\
\nonterm{ConnectorAnn} & \defeq & (\literal{Markers} \alt \typeval{None} \sep \literal{SVGShapePath} \sep \literal{linear} \alt \literal{catmull-rom} \alt \cdots{}) \\
\nonterm{IndicatorAnn} & \defeq & (\literal{line} \alt \literal{area} \alt \literal{arrow} \alt \cdots{} \sep \typeval{Expression})
\end{array} \\\\[-1.2em]
\multicolumn{1}{l}{\underline{\textbf{\textit{Placement}}}} \\
\begin{array}{lcl}
\nonterm{Position} & \defeq & \nonterm{FixedPos} \alt \nonterm{Anchor1D} \alt \nonterm{Anchor2D} \\
\nonterm{Anchor1D} & \defeq & \literal{auto} \alt \literal{start} \alt \literal{mid} \alt \literal{end}, 
\nonterm{Anchor2D}  \defeq \literal{auto} \alt \literal{upLeft} \alt \literal{midRight} \alt \cdots{} \\
\nonterm{FixedPos} & \defeq & \{\dictkey{type} \defeq{} \literal{data} \alt \literal{pixel} \sep \dictkey{x} \defeq{} \nonterm{Coord} \sep \dictkey{y} \defeq{} \nonterm{Coord}\} \\
\nonterm{Coord} & \defeq & \typeval{string} \alt \typeval{number}
\end{array} 
\end{array}
$%
}
\vspace{-0.25em}
\caption{AnnoGram introduces a top-level primitive to declaratively define annotation targets, types, and placement.}
\label{fig:grammar}
\end{figure}

Given these complications, we choose to frame our extension in the manner of CSS via a \textit{target-and-effect} style.
Each annotation has a \texttt{target} and associated visual effects (each with its own positioning and styling elements). For instance, an annotation might target a specific bar in a bar chart, attach an arrow to it, and then a textual explainer might note that it is an outlier. 
This model aligns with a likely common authoring thought-process: first, an annotation author selects target elements or data points to annotate; next, they choose the type of annotation to apply (e.g., text, enclosure, or indicator); finally, they may optionally customize the annotation with styling, positioning, or connectors. 
We reify this process as collections of annotations that select \texttt{targets}, categorizing \texttt{annotation types}, determining \texttt{placement}, and specifying annotation styles. This declarative design supports integration into GoG, but is more closely aligned with our proof-of-concept implementation and design choices made in Cicero~\cite{kim2022cicero}.

We developed AnnoGram through an iterative process of analyzing common annotation practices~\cite{rahman2024qualitative} and encoding them as first-class constructs in a declarative grammar. Our goal was to align this grammar closely with the needs of implementation, ensuring that it is expressive enough for diverse annotation types yet tractable within for typical rendering pipelines. The remainder of this section describes the high-level elements of our design.

\begin{figure}[t]
    \centering
    \includegraphics[width=\linewidth]{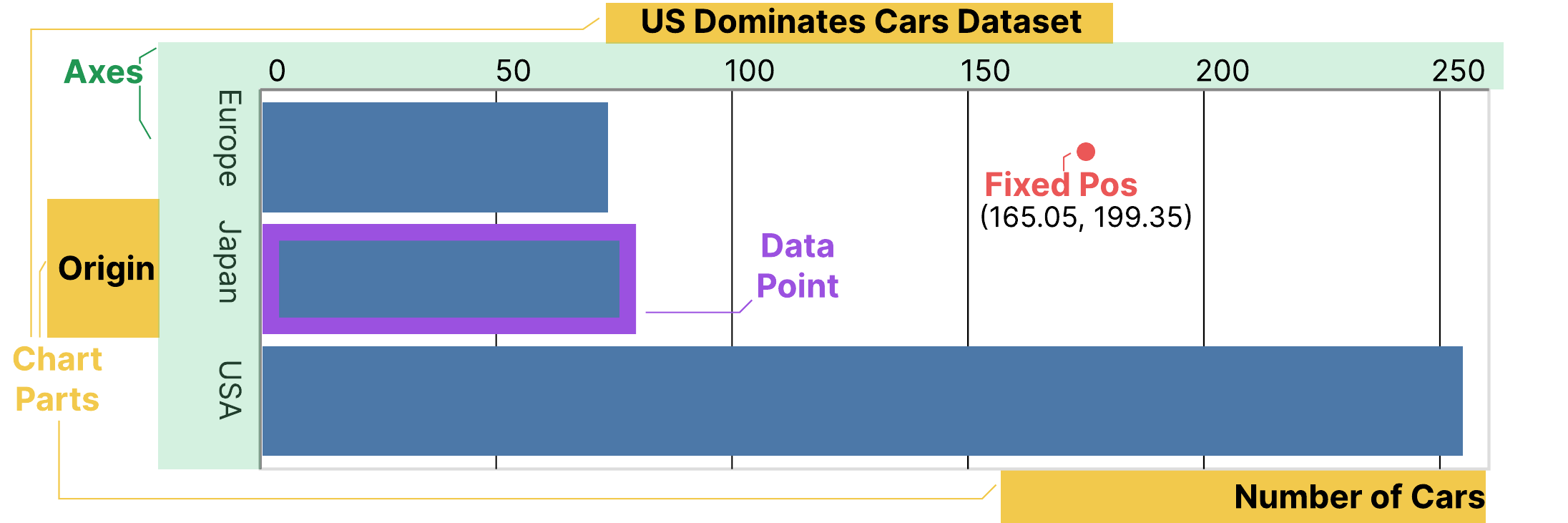}
            \vspace{-1.5em}
    \caption{Supported annotation targets.
    }
    \label{fig:chart-part}
\end{figure}

\parahead{\textit{Target}}
An annotation must first define its \texttt{target}: the specific visualization element it applies to, of which there are several types (see~\autoref{fig:chart-part}). 
First are \textbf{DataPoint}s, which enable annotations to highlight individual data points or subsets defined using direct indexing or expressions. For example, an annotation can highlight all values exceeding a temperature threshold, updating dynamically as the data changes. Beyond data points, annotations can target other key components of a chart. \textbf{Axis} targets allow annotations to be applied to the x- or y-axis, marking significant values, ranges, or structural elements such as tick marks, gridlines, and labels. For instance, an annotation might highlight a time range on the x-axis to provide historical context. \textbf{ChartPart} targets are situating elements such as titles, legends, and axis labels, supporting meta-level explanations such as clarifying what colors or shapes represent in a legend. 
In some cases, annotations are not tied to any specific chart element (\ie{} \textit{None}) and instead serve as standalone notes, positioned independently to offer supplementary context.

\definecolor{dpColor}{HTML}{9A50DF}
\definecolor{axColor}{HTML}{2A9959}
\definecolor{cpColor}{HTML}{F2CB57}
\definecolor{fpColor}{HTML}{EB5656}

Annotations can also interact with each other to create structured, layered explanations. For example, a shaded region might group a set of chart elements, while a nearby explanatory label provides a description of the group's significance. To formalize these relationships, we define two mechanisms: \texttt{Reference} and \texttt{Composite}. In a \texttt{Reference} relationship, one annotation explicitly links to another via its \texttt{id}---for example, a marker denoting an anomaly might reference a separate annotation that explains its cause. \texttt{Composite} annotations, in contrast, group multiple annotations together to describe repeated structures or patterns in the data---such as highlighting seasonal trends across several time intervals using coordinated shapes, markers, and labels. Both of these relationships reflect the concept of annotation ensembles~\cite{rahman2024qualitative, rahman2023exploring, rahman2022qualitative}, in which multiple annotations are intentionally combined to convey complex meanings that single annotations cannot express on their own.

\parahead{\textit{Annotation Types}} Once a target has been defined, the next step is to determine how it should be annotated. Should the annotation appear as a text label, a visual enclosure, or some other graphical element? 
We support a broad set of annotation types based on Rahman \etals{}~\cite{rahman2024qualitative} taxonomy. 
Our grammar accommodates all seven types described in their study, but we focus here on those used in our proof-of-concept implementation. \textbf{Text} annotations convey explanatory context for specific data points, chart regions, or encodings (\eg{} text descriptions in~\autoref{fig:teaser}A). They are highly customizable, with configurable properties such as font size, color, and alignment. \textbf{Enclosure} annotations draw attention to clusters or regions of interest using shapes such as rectangles, ellipses, brackets, or curly braces (e.g., gray highlighted area in~\autoref{fig:teaser}A). These annotations support adjustable padding and stroke styles, offering flexibility in visually presented grouped elements. \textbf{Indicator} annotations mark thresholds, trends, or events through visual markers such as lines, arrows, or highlights (e.g., the arrow showing upward trend of profit in the waterfall chart in~\autoref{fig:teaser}C). Styling options include stroke color, width, and arrowheads. 
\textbf{Connector} annotations link visual elements (\eg{} data points and explanations) using paths or lines (e.g., arrows connecting text descriptions with points of interest in~\autoref{fig:teaser}A). They support various interpolation modes, including \texttt{linear}, \texttt{curved}, and \texttt{stepwise} with other styles.

\parahead{\textit{Placement}} The last key consideration is placement. While \texttt{target} implies constraints---annotations should remain visually associated with their referent---additional positioning strategies offer greater flexibility.
We include a set of placement defaults that users can override, similar to how inline CSS can take precedence over external stylesheets (see \autoref{fig:annotation_pipeline}.3). 
We include two primary positioning modes. \texttt{Fixed Positioning} places annotations in either \texttt{Data-Space} or \texttt{Pixel-Space}. The former anchors annotations relative to encoding scales, enabling reuse across chart types; the latter offers absolute control via screen coordinates. Fine-grained adjustments are supported via \texttt{dx}/\texttt{dy} offsets (not shown in \autoref{fig:grammar}).
Alternatively, \texttt{Relative Positioning \& Anchoring} aligns annotations dynamically with their targets. Anchors may apply in one or both dimensions, snapping to element edges or centers. 
Importantly, manual positioning is never required as annotations receive meaningful placement automatically based on target and layout context, but authors retain full control to specify or refine positions when desired. 
This flexibility increases the utility of automated strategies, ensuring authors can rely on default behaviors while selectively applying manual overrides where precision or customization is needed. 
This dynamic alignment ensures that annotations remain meaningfully positioned as chart dimensions or layouts shift.

\newcommand{\component}[1]{\textbf{#1}}

\section{Proof-of-Concept Implementation}
\label{sec:implimentation}
We developed a proof-of-concept demonstration of AnnoGram as a Vega-Lite extension to explore the utility of our design.

\begin{figure}[t]
    \centering
    \includegraphics[width=\linewidth]{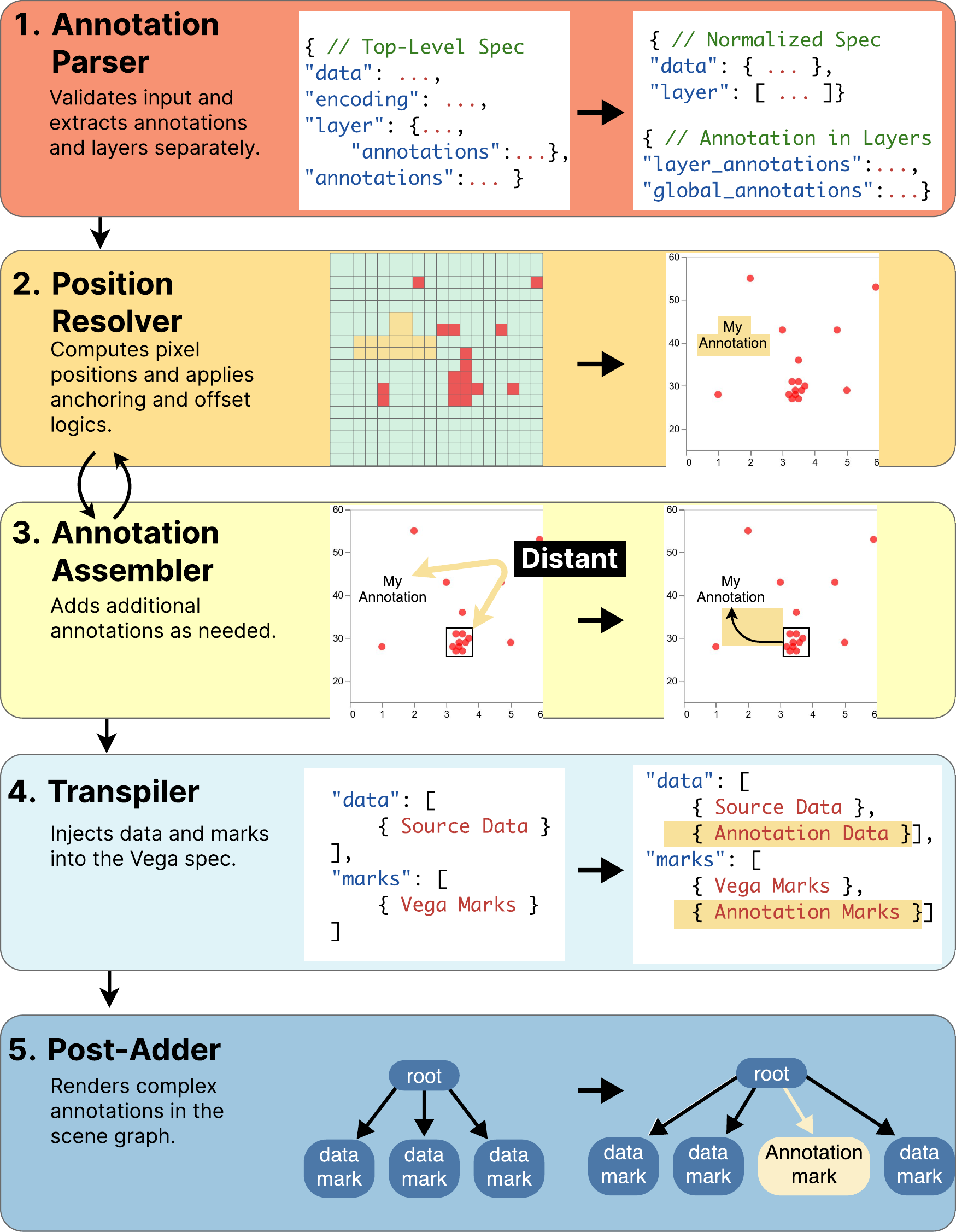}
    \caption{
    Workflow overview of AnnoGram implementation. 
    }
    \label{fig:annotation_pipeline}
\end{figure}

\parahead{Architecture}
Our proof-of-concept extends Vega-Lite’s compilation pipeline to integrate annotations into the specification and rendering process. Annotations require coordination between semantic targets (e.g., data points, axes, chart parts) and their placement in the rendered scene. 
The Vega-Lite pipeline, including normalization, scene graph generation, and rendering, is executed in parallel to extract styles and coordinates, which inform default styling and placement for coherent annotation design with minimal user input.
As in~\autoref{fig:annotation_pipeline}, our library processes annotations in several stages: a \component{Parser}, a \component{Position Resolver}, and an \component{Assembler}, followed by rendering via a \component{Transpiler} and \component{Post-Adder}.

The \component{Parser} takes the input JSON specification, validates it, extracts annotations, and resolves syntax, scales, and default values from the normalized Vega-Lite and Vega specifications. 
The \component{Position Resolver} then analyzes the scene graph to determine optimal annotation placement. User-defined positions (e.g., \texttt{FixedPos}) are prioritized; otherwise, it heuristically selects available space to minimize occlusion (echoing Vega-Label~\cite{vega-label}'s placement strategy). 
For example, in the scatterplot shown in \cref{fig:annotation_pipeline}.2, it identifies  available regions (green) and places annotations in unoccupied areas (yellow) to avoid overlap while accounting for anchoring and offset logic. 
The \component{Assembler} links related marks as needed (\eg{} drawing connectors between labels and enclosures) and iteratively resolves circular or nested references. 
This coordination with the \component{Position Resolver} ensures spatial and semantic coherence across annotation elements. 

Next, the \component{Transpiler} converts annotations into Vega mark encodings using graphical primitives (\eg{} paths, shapes, and text). 
These are inserted into the Vega specification’s \texttt{marks} and \texttt{data} arrays alongside standard chart elements. 
Finally, the \component{Post-Adder} handles annotations requiring advanced rendering, such as non-data-space placement or complex shapes. Elements such as ellipse-shaped enclosures are directly injected into the Vega scene graph with absolute pixel coordinates.

\parahead{Data Dependent Transferable Annotations}
A key contribution of our annotation grammar is support for annotation portability---annotations that remain valid and well-positioned as data or chart types change. This portability is achieved through a combination of user-defined and automatic positioning strategies that decouple annotations from specific mark or layout configurations.

For user-defined positioning, the Annotation \component{Parser} handles both pixel-space and data-space targets. Pixel-space targets are resolved via the scene graph for exact placement, whereas data-space targets use encoding scales extracted from the Vega specification. The \component{Position Resolver} computes annotation positions based on target type, specified positions, 
10 mark types, and chart layout. 
The \component{Position Resolver} and Annotation \component{Assembler} iteratively assign positions by identifying unoccupied regions and minimizing occlusion via a backtracking algorithm. 
During the positioning, relative adjustments using offsets and anchor values (\texttt{Anchor1D}, \texttt{Anchor2D}) are also applied on the data between \component{Position Resolver} and \component{Assembler}. 
This differs from Vega-label's~\cite{vega-label,kittivorawong2024legiblelabellayoutdata} approach, which is limited to repositioning only text annotations by adjusting text mark placements to avoid collisions. 
Whereas our approach supports multiple annotation types (text, enclosure, connector) with flexible, data- and pixel-based positioning across chart types.

\begin{figure}[!t]
    \centering
    \includegraphics[width=\linewidth]{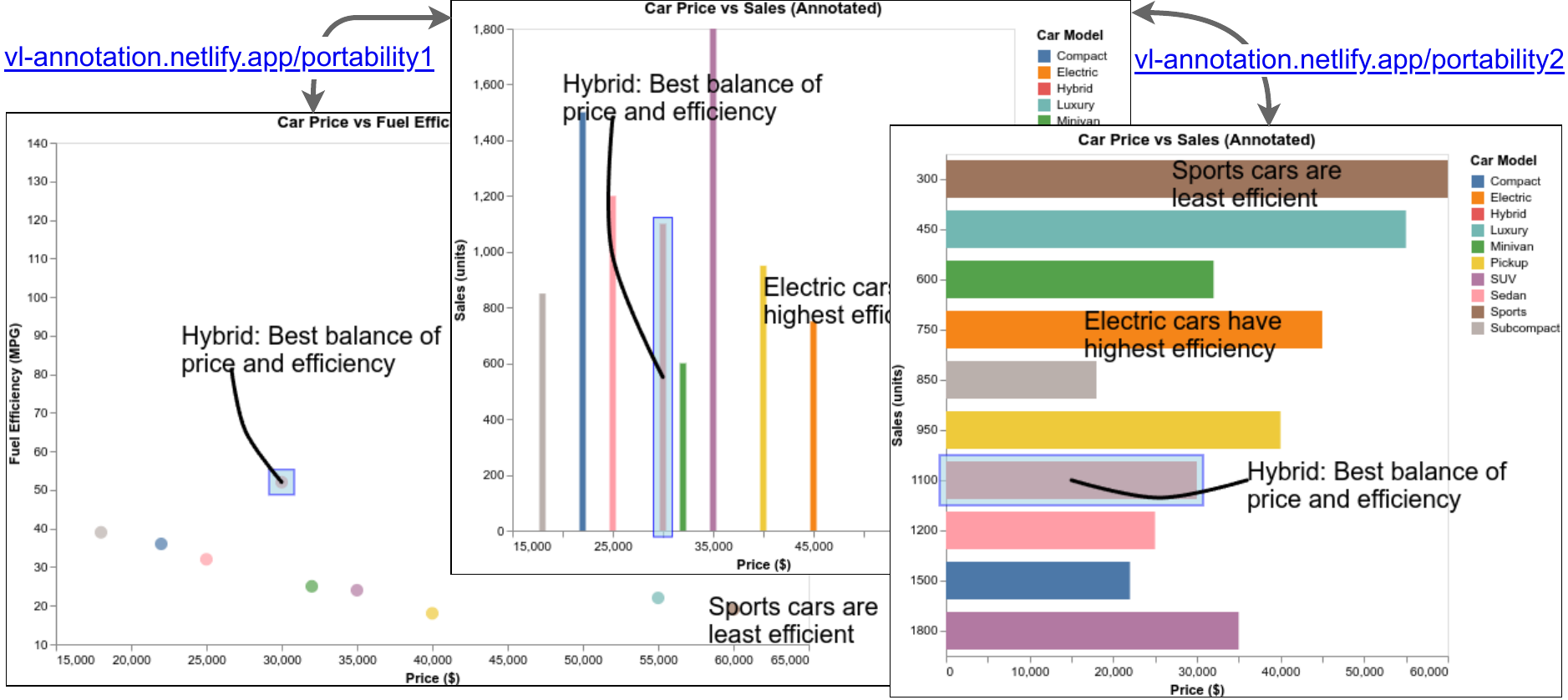}
    \caption{Same annotation specification applied to three different chart types with different fields of the same data. AnnoGram dynamically detects the target mark types and adds annotations. 
    }
    \label{fig:portability}
    \vspace{-1em}
\end{figure}

These design choices enable portability by resolving annotation placement at runtime based on target semantics, chart encodings, and available space, making annotations robust to changes in chart structure, reducing the need for manual adjustments during design. 
For example, a text annotation targeting a data point in a scatterplot can be reused in a bar chart of the same data without any changes. In \autoref{fig:portability}, a single annotation specification results in three different, context-appropriate placements, depending on the target and chart configuration.

\section{Evaluation}
We conducted a heuristic comparative evaluation of our proof-of-concept implementation against eight existing tools.
We do so by considering the differences between annotated bar, line, and scatterplots produced using each tool.
Evaluation focused on annotation-specific and total lines of code (LOC) and four qualitative dimensions: \textbf{1st-class annotation support}, which assesses whether annotations are structured components within the visualization model rather than external overlays; 
\textbf{intuitiveness}, which measures how naturally the annotation syntax reflects the data and chart structure, and how easily common annotation tasks can be expressed; 
\textbf{error proneness}~\cite{green1996usability,cdn2}, which captures the likelihood of mistakes during annotation authoring or editing, such as misalignment or broken references; 
and \textbf{portability}, which examines the ability of annotations to adapt across different datasets or chart types without manual rework. 
\autoref{tab:annotation_comparison} summarizes the findings; examples and materials are available at \asLink{https://github.com/shape-vis/codelines-comparison}{github.com/shape-vis/codelines-comparison}.

Across the evaluated tools, the degree of semantic integration was the primary factor shaping authoring complexity, intuitiveness, error susceptibility, and portability. 
General-purpose charting tools such as D3, Vega, and Vega-Lite lack built-in annotation constructs, requiring manual positioning, anchoring, and updates, resulting in high annotation LOC, cognitive load, and user error, with ``Hard'' ratings for intuitiveness and portability, and ``High'' error proneness.
Whereas d3-annotate, ggplot2, and ggplot2-annotate introduce annotation primitives that reduce LOC and initial authoring effort, but weak semantic coupling still requires manual adjustments during updates or context changes. 
As reflected in their ``Medium'' ratings for intuitiveness and error proneness, these libraries partially improve authoring efficiency but do not fully address maintenance or dynamic adaptation challenges. 
HighCharts simplifies initial authoring but retains similar limitations, requiring user-driven updates for structural changes. 
Visual editors such as PowerPoint or Figma achieve ``Easy'' ratings for intuitiveness via direct manipulation, but disconnect annotations from data and chart structure, making them static and poorly portable across datasets or designs. 

In contrast, AnnoGram embeds annotations as structured, semantically aware visualization components. Annotations are declaratively attached to data points, axes, and chart elements, enabling automated placement, anchoring, and runtime adaptation. This design substantially reduces annotation LOC compared to low-level libraries, minimizes manual corrections during data or layout changes, and results in consistently ``Easy'' ratings across all evaluated dimensions. 
Updates to the visualization automatically propagate to annotations, lowering error-proneness and enhancing portability without requiring reauthoring. 
Overall, the evaluation highlights a consistent pattern: tools without semantic annotation integration impose high authoring and maintenance overhead and are prone to errors during evolution. Systems that partially address authoring burden without semantic binding improve initial usability but struggle with long-term robustness. 
Our AnnoGram implementation demonstrates that embedding structured annotation semantics within the visualization grammar substantially improves authoring efficiency, reduces error rates, and supports scalable, maintainable annotation workflows across different datasets and visualization types.
Additional consideration of the fine-grain differences between these notations is worthwhile future work, and may be aided by Kruchten et al.’s multi-notation gallery approach~\cite{kruchten2023metrics}.

\newcommand*\tabRot{\rotatebox{90}}
\newcommand{\hardCell}{Hard} %
\newcommand{\medCell}{\cellcolor{gray!15}Med.} %
\newcommand{\easyCell}{\cellcolor{cyan!25}Easy} %

\newcommand{\highCell}{High}
\newcommand{\midCell}{\cellcolor{gray!15}Med.}
\newcommand{\lowCell}{\cellcolor{cyan!25}Low}

\newcommand{\myCheck}{\cellcolor{cyan!25}$\checkmark$}
\newcommand{\myX}{\texttimes{}}

\newcommand{\highPortCell}{\cellcolor{cyan!25}Easy}
\newcommand{\midPortCell}{\cellcolor{gray!15}Med.}
\newcommand{\lowPortCell}{Hard}

\begin{table}[t!]
    \caption{
        We implemented a trio of example charts 
        with annotations across eight other tools. We give lines of code required for \emph{just} the annotation (Ann.), as well as coding of several qualitative properties.
    }
    \vspace{-0.25em}
    \centering
    \setlength{\tabcolsep}{2.4pt} %
    \resizebox{\linewidth}{!}{%
    \begin{tabular}{l|l|rrrrrr|cccc|}

        \rotatebox[origin=bl]{90}{Programmability} & Tool & 
        \rotatebox[origin=bl]{90}{Bar Ann. LOC}\hspace{4pt}  & 
        \rotatebox[origin=bl]{90}{Bar LOC}\hspace{4pt} & 
        \rotatebox[origin=bl]{90}{Line Ann. LOC}\hspace{4pt} & 
        \rotatebox[origin=bl]{90}{Line LOC}\hspace{4pt} & 
        \rotatebox[origin=bl]{90}{Scatter Ann. LOC}\hspace{4pt} & 
        \rotatebox[origin=bl]{90}{Scatter LOC}\hspace{4pt} & 
        \rotatebox[origin=bl]{90}{1st-class Support} & 
        \rotatebox[origin=bl]{90}{\hspace{-1pt}Intuitiveness} & 
        \rotatebox[origin=bl]{90}{\hspace{-1pt}Error Proneness} & 
        \rotatebox[origin=bl]{90}{\hspace{1pt}Portability} \\

        \arrayrulecolor{gray}
        \specialrule{.4pt}{1pt}{1pt}
        \arrayrulecolor{black}

        \multirow{4}{*}{\rotatebox[origin=c]{90}{High}} & d3~\cite{bostock2011d3}            & 151 & 81  & 175 & 76  & 170 & 104 & \myX{}     & \hardCell{} & \highCell{} & \hardCell \\
                                      & d3-annotate~\cite{lu_d3-annotation} & 161 & 81  & 174 & 77  & 177 & 99  & \myCheck{} & \medCell{}  & \midCell{}  & \midPortCell \\
                                      & ggplot2~\cite{wickham2011ggplot2}   & 120 & 22  & 128 & 43  & 208 & 101 & \myCheck{} & \medCell{}  & \highCell{} & \midPortCell \\
                                      & ggplot2-annotate~\cite{wickham2011ggplot2} & 73 & 20  & 117 & 42  & 178 & 101 & \myCheck{} & \medCell{}  & \highCell{} & \midPortCell \\

        \arrayrulecolor{gray}
        \specialrule{.4pt}{1pt}{1pt}
        \arrayrulecolor{black}

        \multirow{4}{*}{\rotatebox[origin=c]{90}{Low}} & HighCharts~\cite{highcharts_docs}   & 82  & 32  & 231 & 82  & 169 & 79  & \myCheck{} & \medCell{}  & \lowCell{}  & \hardCell \\
                                      & Vega~\cite{satyanarayan2015reactive} & 515 & 161 & 491 & 219 & 387 & 186 & \myX{}     & \hardCell{} & \highCell{} & \hardCell \\
                                      & Vega-Lite~\cite{satyanarayan2016vega} & 259 & 23  & 253 & 29  & 177 & 34  & \myX{}     & \hardCell{} & \highCell{} & \hardCell \\
                                      & \textbf{VL Annotation} & \textbf{101} & \textbf{19} & \textbf{95} & \textbf{25} & \textbf{79} & \textbf{31} & \textbf{\myCheck} & \textbf{\easyCell} & \textbf{\lowCell} & \textbf{\easyCell} \\

        \arrayrulecolor{gray}
        \specialrule{.4pt}{1pt}{1pt}
        \arrayrulecolor{black}

        \multirow{2}{*}{--} & Visual Editors & \multirow{2}{*}{\ \ --} & \multirow{2}{*}{\ \ --} & \multirow{2}{*}{\ \ --} & \multirow{2}{*}{\ \ --} & \multirow{2}{*}{\ \ --} & \multirow{2}{*}{\ \ --} & \multirow{2}{*}{--} & \cellcolor{cyan!25} & \cellcolor{cyan!25} & \multirow{2}{*}{Hard} \\
         & \ \ {\small (PowerPoint, Figma, etc.)} &  &  &  &  &  &  &  & \cellcolor{cyan!25}\multirow{-2}{*}{Easy} & \cellcolor{cyan!25}\multirow{-2}{*}{Low} & \\

        \arrayrulecolor{gray}
        \specialrule{.6pt}{1pt}{1pt}
        \arrayrulecolor{black}
    \end{tabular}}
    \label{tab:annotation_comparison}
    \vspace{-1.5em} %
\end{table}

\section{Discussion} 

We present AnnoGram, an extension to GoG that treats annotations as first-class elements alongside encodings, marks, and layout, demonstrated via a proof-of-concept implementation as a Vega-Lite extension. Its declarative design enables runtime positioning and rendering, avoiding manual coordination. By referencing semantic targets and using scale-aware positioning, annotations remain portable across chart types and layouts, improving reusability, reducing effort, and minimizing errors compared to systems lacking integrated support (see \autoref{tab:annotation_comparison}).

Our proof-of-concept implementation realizes a subset of AnnoGram's annotation capabilities. 
To ensure compatibility and ease of deployment, it is implemented as an external wrapper around Vega-Lite rather than modifying its compiler directly. 
This architectural choice simplifies integration but limits access to internal optimization and tooling infrastructure. In its current form, the library restricts each target to a single annotation per type, which reduces flexibility in scenarios that demand layered or redundant annotations. 
It also currently does not support compositions such as facet, layer, or concat, though these are planned for future versions. While some basic aggregations work in the current implementation, complex aggregations involving multiple mark types are not fully tested and may cause failures.
Additionally, it supports only the most common annotation types observed in prior works~\cite{rahman2024qualitative}, excluding less frequent or interactive forms. These decisions reduce complexity but constrain extensibility, especially for use cases needing richer semantics or native tooling support.

Although grounded in prior analysis~\cite{rahman2024qualitative}, AnnoGram reflects one modeling approach that emphasizes simplicity and declarative clarity. Its current flat structure streamlines authoring and parsing, but limits expressiveness for nested or highly coordinated ensembles. Furthermore, our evaluation, based on a small set of expert-authored examples, also limits generalizability. 
Future work should explore alternative structures, broader usage contexts, and integration with recommendation systems and authoring tools. 
AnnoGram's declarative structure opens promising directions for integration with annotation recommendation systems and tooling support. 
Its alignment with GoG concepts also suggests potential extensions to other under-supported constructs, such as explanatory annotations, visual embellishments, or accessibility layers. Embedding AnnoGram into grammar compilers could further enable native optimization, richer design semantics, and unified authoring environments.

\setstretch{0.948}
\acknowledgments{
  The authors wish to thank our anonymous reviewers. This work was supported in part by
  NSF award 2402719. The first two authors contributed equally to this work.}

\bibliographystyle{abbrv-doi-hyperref}
\bibliography{main}

\end{document}